\documentclass[aps,pra,twocolumn,groupedaddress,showpacs,superscriptaddress,amssymb,amsmath]{revtex4-2}
\usepackage{natbib}
\usepackage[utf8]{inputenc}
\usepackage{graphicx}
\usepackage{tabularx}
\usepackage{xcolor}
\usepackage{amsmath}

\usepackage{comment}
\usepackage{dcolumn}
\usepackage{hyperref}
\hypersetup{
	colorlinks=true,
	linkcolor=blue,
	citecolor=blue
}
\usepackage{bm}
\usepackage{epsf}
\usepackage{braket}
\usepackage{tensor}
\usepackage{soul}
\usepackage{mathrsfs}
\usepackage{mathtools}
\usepackage{multirow}

\newcommand{\mybra}[1]{\langle #1 |}
\newcommand{\myket}[1]{| #1 \rangle}

\usepackage{amsfonts}

\newcommand{\be}{\begin{equation}}
	\newcommand{\ee}{\end{equation}}
\newcommand{\bea}{\begin{eqnarray}}
	\newcommand{\eea}{\end{eqnarray}}

\makeatletter
\newsavebox{\@brx}
\newcommand{\llangle}[1][]{\savebox{\@brx}{\(\m@th{#1\langle}\)}%
	\mathopen{\copy\@brx\kern-0.5\wd\@brx\usebox{\@brx}}}
\newcommand{\rrangle}[1][]{\savebox{\@brx}{\(\m@th{#1\rangle}\)}%
	\mathclose{\copy\@brx\kern-0.5\wd\@brx\usebox{\@brx}}}
\makeatother

\begin{document}


\title{Emergence of distinct relaxation behaviour and Quantum Regression Theorem in the Ultra-strong Coupling Limit}
\author{Sakil Khan}
\email{sakil.khan@students.iiserpune.ac.in}
\affiliation{Department of Physics,
		Indian Institute of Science Education and Research, Pune 411008, India}
\author{Bijay Kumar Agarwalla}
\email{bijay@iiserpune.ac.in}
\affiliation{Department of Physics,
		Indian Institute of Science Education and Research, Pune 411008, India}
\date{\today}
\begin{abstract}
In the framework of open quantum systems, we derive the dynamical equation governing two-time correlation functions in the ultra-strong coupling (USC) regime between the system and its environment. Unlike the case of the standard weak-coupling regime, in the USC case, we find distinct relaxation behavior for two-time correlators depending on the types of the operators involved in the correlation function. 
Interestingly, the Quantum Regression Theorem (QRT) emerges after the fastest relaxation time-scale, which is governed by the system-bath coupling strength. We exemplify our findings for the dissipative spin-boson model and further find excellent agreement with the numerically exact hierarchical equations of motion (HEOM) method.
\end{abstract}
  \maketitle  

{\it Introduction.--}
Understanding the quantum dynamics and long-time behavior of a system of interest coupled to a thermal environment remains a central topic in open quantum system research \cite{breuer,Carmichael,schaller2014open,xiong2020exact,PhysRevB.107.125149,PhysRevE.100.022111,PhysRevB.97.134301,PhysRevB.86.155424,PhysRevLett.121.170402,Gabriel_Review_NBD,PRXQuantum.5.020201}. In this context, it is not only crucial to analyze the evolution of the system’s density matrix but also the dynamics of multi-time correlation functions \cite{breuer,Carmichael,Gribben2022usingenvironmentto,PhysRevResearch.5.033078}. These correlation functions serve as key dynamical quantities, often directly linked to experimentally accessible observables, such as fluorescence and absorption spectra in molecular spectroscopy \cite{PhysRevA.107.023717,Bogaczewicz_2023}, photon bunching and anti-bunching in quantum optics \cite{PhysRevLett.39.691,PhysRevLett.131.143601,PhysRevA.108.023727}, etc. Compared to the reduced density matrix, these functions can often offer deeper insights into both the dynamical and steady-state properties of the system \cite{breuer,Carmichael,doi:10.1143/JPSJ.12.570,RevModPhys.81.1665,Sieberer_2016}. 

The dynamics of the reduced density matrix is conventionally studied using the quantum master equations (QMEs), such as the Redfield and Lindblad formulations \cite{Carmichael,breuer,Alicki,davies,REDFIELD19651,rivas2012open,lidar2019lecture,GORINI1978149,Lindblad,Gorini}. In contrast, the multi-time correlation functions are traditionally investigated using the Quantum Regression Theorem (QRT) \cite{Carmichael,breuer, Sakil1,qrt,otoc1,qn,Khan_2024}. However, both the QME and QRT frameworks are primarily restricted to the regime of weak system-environment coupling.

In recent years, there has been a growing interest in understanding the effects of {\it strong} system-environment coupling. This interest is driven both by fundamental questions in strong-coupling quantum thermodynamics \cite{roadmap, PhysRevB.84.161414,PhysRevLett.116.020601,RevModPhys.92.041002,PhysRevLett.102.210401,PhysRevE.95.032139,PhysRevLett.124.160601,PhysRevLett.120.120602,PhysRevB.107.195117, PhysRevX.10.031040,PhysRevLett.127.250601,PhysRevResearch.4.043168,Vinjanampathy01102016,10.1116/5.0073853,PhysRevE.102.012155,PhysRevA.111.032214} and by the prospect of numerous important applications that include quantum thermometry \cite{PhysRevA.96.062103,PhysRevB.98.045101}, quantum thermal machines \cite{PhysRevE.95.032139,s2025}, or other emerging quantum technologies that operate beyond the weak-coupling regime \cite{Wang2015,Mitchison_2015,PhysRevA.84.012103,PhysRevLett.116.120801,PhysRevLett.109.233601}. A substantial amount of research in this domain has focused on analyzing the dynamics of the reduced density matrix using various analytical and numerical techniques, such as the reaction coordinate method \cite{PhysRevA.104.052617,PRXQuantum.4.020307,PhysRevA.90.032114,PhysRevE.105.024126}, polaron transformation approach \cite{PRXQuantum.4.020307,PhysRevB.101.235114,PhysRevB.110.174304}, the hierarchical equations of motion (HEOM) approach \cite{PhysRevA.97.033411,Mangaud_2018,doi:10.1143/JPSJ.58.101,PhysRevB.94.201407}, and the time-evolving matrix product operator (TEMPO) algorithm \cite{Strathearn2018,PhysRevLett.126.200401,Fux2022,PhysRevResearch.5.033078,modi_process_tensor}. Notably, a recent study \cite{PhysRevA.106.042209} in this direction introduced a novel perturbative method specifically designed to address reduced dynamics in the {\it ultra}-strong coupling (USC) regime. However, with only a few exceptions \cite{CALDEIRA1983587},  the analysis of multi-time correlation functions beyond the standard weak-coupling limit remains largely unexplored in the literature. This is the gap we aim to address in this letter.

In this work, we investigate  the dynamical properties of two-time correlation functions in the USC regime.  We employ the Nakajima-Zwanzig projection operator formalism \cite{breuer} to analyze the dynamics of the correlation functions in the USC regime \cite{10.1116/5.0073853,PhysRevLett.127.250601}.
As mentioned earlier, in the context of the weak-coupling framework, the QRT serves as a standard tool for computing correlation functions.  A natural and important question arises: Does the QRT remain valid in the USC regime? Interestingly, we find that in the USC regime, depending on the type of two-time correlator, there could be two distinct time-scales and the QRT emerges once the fastest time-scale has elapsed. We provide here a general analytical framework for evaluating two-time correlation functions in the USC regime. This fully analytical treatment offers deeper insights into the behavior of correlation functions. Remarkably, our analysis uncovers a novel feature -- the emergence of two distinct relaxation time scales in the two-time correlation functions. 
Interestingly, this behavior does not appear in the weak-coupling regime and is a distinctive feature of the strong system-bath coupling regime. We validate  our findings against results obtained from the numerically exact hierarchical equations of motion (HEOM) method, observing excellent agreement.

{\it Setup and two-time correlation functions.--}
We begin by introducing the model of interest and outline the perturbative scheme employed in the USC regime. The total Hamiltonian of the setup can be decomposed as 
$H = H_{S}+H_{R}+H_{\rm SR}$,
where 
$H_{S}$ denotes the Hamiltonian of the system of interest, $H_{R}$ corresponds to the reservoir (bath) Hamiltonian, and $H_{\rm SR}$ captures the interaction between the system and the reservoir. In this work, we consider a finite-dimensional quantum system coupled to a bosonic thermal reservoir. The total Hamiltonian is explicitly written as 
\begin{eqnarray} 
\label{eq:Ham2}
H &= &H_{S}+\sum_{k} \Omega_{k} b_{k}^\dagger b_{k}+ S \otimes\sum_{k} \alpha_{k} (  b_{k}^\dagger+ b_{k}),
\end{eqnarray}
where  $b_{k}$ ($b^\dagger_{k}$) denote the bosonic annihilation (creation) operator for the 
$k$-th mode. The third term in Eq.~\eqref{eq:Ham2} represents the system-bath interaction, where a generic hermitian system operator $S$ couples to the $k$-th bath mode with coupling strength 
$\alpha_{k}$. We now assume that the set
$\{\myket{n}\}$ forms the eigenbasis of the system operator 
$S$, such that 
$S=\sum_{n} \theta_{n}\myket{n}\mybra{n}$, where $\theta_n$ are the corresponding non-degenerate eigenvalues. In this basis, the system part of the total Hamiltonian can be expressed as
 \begin{align}
 \label{eq:Ham_basis}
     H_{S}=\sum_{n} \epsilon_{n}\myket{n} \mybra{n}+\sum_{n\neq m} J_{nm}\myket{n}\mybra{m},
 \end{align}
where the diagonal terms $\epsilon_{n}$ are real and the off-diagonal elements satisfy $J^*_{nm}=J_{mn}$, ensuring hermiticity. Based on this decomposition, we partition the total Hamiltonian into an unperturbed part and a perturbation as $H=H_{0}+\lambda V$, where 
\begin{eqnarray} 
\label{eq:H0}
H_{0} &= \sum_{n} \epsilon_{n}\myket{n} \mybra{n}+H_{R}+H_{\rm SR}
\end{eqnarray}
and 
\begin{align}
\label{eq:V}
V=\sum_{n\neq m} J_{nm}\myket{n}\mybra{m}.
\end{align}
The parameter $\lambda$ is a dimensionless bookkeeping parameter introduced to systematically track the perturbative expansion in terms of $V$.  It is important to emphasize that within the standard weak-coupling framework, the system-reservoir interaction  
$H_{\rm SR}$ is typically treated as a small perturbation. In contrast, in the USC regime, the off-diagonal part of the system Hamiltonian $V$ in Eq.~\eqref{eq:V} which governs coherent transition between the eigenstates of the operator $S$ is treated perturbatively, while $H_0$ in Eq.~\eqref{eq:H0} that includes $H_{\rm SR}$
is incorporated into the unperturbed Hamiltonian and is not assumed to be weak. 

We rewrite the unperturbed Hamiltonian $H_{0}$ as
\begin{align}
    H_{0}=\sum_{n} \tilde{\epsilon}_{n}\myket{n}\mybra{n}+\sum_{n,k} \Omega_{k} b^{n \dagger}_{k} b^{n}_{k},
    \label{H0-dressed}
\end{align}
where we define the displaced bath annihilation operator as $b^{n}_{k}=b_{k}+\frac{\theta_{n}\alpha_{k}}{\Omega_{k}}\myket{n} \mybra{n}$ and the dressed onsite energy as $ \tilde{\epsilon}_{n}=\epsilon_{n}-\sum_{k}\frac{\alpha_{k}^{2}}{\Omega_{k}}\theta_{n}^{2}=\epsilon_{n}-\delta\epsilon_{n}$. 
Note that, the reorganization energy can be expressed as $\delta\epsilon_{n}=\int d\Omega \frac{J(\Omega)\theta_{n}^{2}}{\Omega
}$, where the bath spectral density is given by  $J(\Omega)= \sum_{k}\alpha_{k}^{2} \delta(\Omega-\Omega_{k})$.
We consider the initial system-bath state to be 
of the form $\rho_{\rm SR}(0)=\sum_{n}p_{n}\myket{n}\mybra{n}\otimes \rho^{n}_{R}$, where $\rho^{n}_{R}$ is the thermal Gibbs state with respect to the displaced bath Hamiltonian, $H^{n}_{R}=\mybra{n}\sum_{k,m} \Omega_{k} b^{m \dagger}_{k} b^{m}_{k} \myket{n}$. We now introduce a projection operator ${\cal P}$, whose action is defined as $ {\cal P} X_{\rm SR}=\sum_{n}\mybra{n} {\rm Tr}_{R}[X_{\rm SR}]\myket{n} \myket{n}\mybra{n}\otimes \rho^{n}_{R}$, and projection operator ${\cal Q}$ is defined as ${\cal Q}=I-{\cal P}$. Note that, 
the initial state is invariant under the action of the projection operator ${\cal P}$, i.e., ${\cal P} \rho_{\rm SR}(0)=\rho_{\rm SR}(0)$, and thus ${\cal Q}\rho_{\rm SR}(0)=0$.
Our primary aim in this work is to calculate the two-point correlation function of the form $C(\tau,t)=\langle O_{1}(t)O_{2}(t+\tau) \rangle$ in the USC limit which we discuss next. 

\vspace{0.3cm}
{\it Correlation function in the USC regime.--} We develop an analytical framework to compute the two-time correlation function in the USC regime.  The two-time correlation function of the form  $C(\tau,t)=\langle O_{1}(t)O_{2}(t+\tau) \rangle$ can be written as 
\begin{align}\label{effd}
   C(\tau,t)=  \langle O_{1}(t)O_{2}(t+\tau) \rangle={\rm Tr}_{\rm S}[O_{2}(0)\xi_{S}(\tau,t)],
 \end{align}
where we introduce the effective density matrix $\xi_{S}(\tau,t)={\rm Tr}_{R}[\xi_{\rm SR}(\tau,t)]={\rm Tr}_{R}[e^{-iH\tau} \rho_{\rm SR}(t)\, O_{1}(0)\,e^{iH\tau}]$.
The full dynamical behavior of the correlation function is captured by the effective density matrix, $\xi_{S}(\tau,t)$. It is important to emphasize that $\xi_{S}(\tau,t)$ is not the actual system density matrix but an effective construction, and thus may not fulfill all the properties of a physical density matrix. Eq.~\eqref{effd} can be rewritten as 
\begin{align}
\label{corr-gen}
    C(\tau,t)=\sum_{n,m} \mybra{m}O_{2}\myket{n}\, \mybra{n}\xi_{S}(\tau,t)\myket{m}.
\end{align}
This indicates that if the operator 
$O_{2}$ is diagonal in the eigenbasis of 
$S$ i.e., $\mybra{m}O_{2}\myket{n}= O_{2m} \delta_{mn}$, then the evaluation of 
$C(\tau,t)$ requires only the diagonal elements of 
$\xi_{S}(\tau,t)$ i.e., $ \mybra{n}\xi_{S}(\tau,t) \myket{n}$. However, if 
$O_{2}$  has off-diagonal components in this basis, the off-diagonal elements of 
$\xi_{S}(\tau,t)$ can also contribute to the correlation function.
We therefore study the dynamics of $\xi_{S}(\tau,t)$ in the USC limit and investigate how QRT emerges. 

In the weak-coupling regime, the QRT is a widely used tool for computing correlation functions \cite{Carmichael,breuer, Sakil1,qrt,otoc1,qn}. Briefly, the QRT states that the knowledge of the time evolution of a single-point function (density matrix) is sufficient to determine the time evolution of two-point correlation functions.
More explicitly, if the density matrix of the system, $\rho_{S} (t)$, evolves under the dynamical map $\Phi(t,0)$ as
$\rho_{S} (t)=\Phi(t,0)\rho_{S} (0)$,
then the QRT for the two-time correlation functions reads as \cite{Carmichael,breuer}
\begin{align}\label{eqqrt}
	\langle O_{1}(t)O_{2}(t+\tau)\rangle&\!=\!{\rm Tr}_{S}\Big[O_{2}\Phi(t+\tau,t)[\xi_{S}(0,t)]\Big].
\end{align}
Note that, to derive the QRT in
Eq.~\eqref{eqqrt}, apart from assuming weak coupling between the
system and bath, the strong Markovian limit, i.e., $ \tau\gg\tau_{B}$ needs to be assumed \cite{Sakil1}, where $\tau_B$ is the characteristic decay time-scale for the bath correlation function. 

We now investigate how and under what conditions QRT emerges in the USC limit. In this context, it is important to highlight that, very recently, in Ref.~\cite{PhysRevA.106.042209}, the author derived a quantum master equation in the USC limit. Specifically, the author showed that the diagonal part of the reduced density matrix of the system $\rho_{S}(t)$ (in the basis states $|n\rangle$ of the coupling operator $S$) is governed by the Pauli-type master equation 
\begin{align}
\label{opd}
   &  \frac{d}{dt}p_{n}(t)=\lambda^2 \sum_{m\neq n}\Big[\gamma_{nm} \, p_{m}(t)-\gamma_{mn} \, p_{n}(t)\Big],
\end{align}
where $p_{n}=\mybra{n}\rho_{S}(t)\myket{n}$ and $\gamma_{nm}$ are rate constants, proportional to the bath correlation function.
Additionally, it was shown that the off-diagonal elements of the density matrix, $\rho^{nm}_{S}=\mybra{n}\rho_{S}(t)\myket{m}$, with $n \neq m$, can be deduced in terms of the diagonal elements $p_{n}(t)$ \cite{PhysRevA.106.042209}. Keeping this in mind, in what follows we will analyze two different scenarios: First, we focus on a situation where the operator $O_1$ is diagonal but $O_2$ may or may not be diagonal. We then generalize our study for non-diagonal $O_1$.  

\noindent {\it Case 1: Diagonal  $O_{1}$ Operator.--}
Let us now consider the first case and assume that the operator $O_{2}$ in Eq.~\eqref{corr-gen} is  also diagonal in the basis of the system operator $S$. Therefore, to capture the dynamics of $\langle O_{1}(t)O_{2}(t+\tau)\rangle$, only the diagonal elements of the effective density matrix $ \xi_{S}(\tau,t)$ are enough. Thus, the statement of QRT [as given in Eq.\eqref{eqqrt}] translates to showing that the diagonal elements of the effective density matrix follow the same equation as the actual density matrix, i.e., Eq.~\eqref{opd}. 

\subsubsection{Dynamical equation for the diagonal elements of the effective density matrix}
Let us denote the diagonal part of $\xi_{S}(\tau,t)$ by $\xi^n_{S}(\tau,t)=\mybra{n}\xi_{S}(\tau,t) \myket{n}$.
To derive the dynamical equation for $\xi^n_{S}(\tau,t)$, we employ the Nakajima-Zwanzig projection operator method \cite{breuer}. 
In the interaction picture with respect to the unperturbed Hamiltonian $H_0$ [Eq.~\eqref{H0-dressed}], $\xi_{\rm SR}(\tau,t)=e^{-iH\tau} \rho_{\rm SR}(t)\, O_{1}(0)\,e^{iH\tau}$
satisfies the following equation
\begin{align}
     \frac{d}{d\tau} {\tilde{\xi}}_{\rm SR}(\tau,t)= {\cal L} (\tau) \, \tilde{\xi}_{\rm SR}(\tau,t),
\end{align}
where the action of the superoperator ${\cal L}(\tau)$ is defined as ${\cal L}(\tau) * =-i[\tilde{V}(\tau), *] $. The  interaction picture operators $\tilde{\xi}_{\rm SR}(\tau,t)$ and $\tilde{V}(\tau)$ are given by 
\begin{align}
    & \tilde{\xi}_{\rm SR}(\tau,t)=e^{iH_{0}\tau}\; \xi_{\rm SR}(\tau,t) \;e^{-iH_{0}\tau},\nonumber\\
     & \tilde{V}(\tau)=e^{iH_{0}\tau}\; V\;e^{-iH_{0}\tau}.
 \end{align}
Employing the Nakajima-Zwanzig projection operator method, one can  show that the dynamics of the effective density matrix is governed by the following equation (See \cite{supp} for the details of the derivation)
\begin{align}\label{eqnzm}
    \frac{d}{d\tau} {\cal P} \, \tilde{\xi}_{\rm SR}(\tau,t)=K(\tau) \, {\cal P} \, \tilde{\xi}_{\rm SR}(\tau,t)+ \textcolor{black}{I(\tau) \,  {\cal Q} \, \tilde{\xi}_{\rm SR}(0,t)},
 \end{align}
The first and second terms of Eq.~\eqref{eqnzm}  represent the homogeneous and inhomogeneous terms, respectively. The homogeneous term 
up to $\mathcal{O}(\lambda^2)$ is given by 
\begin{equation}
\label{eqih}
\!\!K(\tau)\, {\cal P} \tilde{\xi}_{\rm SR}(\tau,t)\!=\!\lambda^2 \int^{\tau}_{0} ds \, {\cal P} \, {\cal L}(\tau) \, {\cal L}(s)\, {\cal P} \, \tilde{\xi}_{\rm SR}(\tau,t).
\end{equation}
Interestingly, this homogeneous term is akin to what we observe in the weak coupling case.
Under the strong Markov approximation, i.e., when $\tau\gg \tau_{B}$, the upper limit of the integration variable in Eq.~\eqref{eqih} can be extended to $\infty$. In a similar way, the inhomogeneous part up to $\mathcal{O}(\lambda^2)$ can be computed and its explicit expression is given as 
\begin{eqnarray}
\label{eqinhaa-main}
&&I (\tau) {\cal Q} \tilde{\xi}_{\rm SR}(0,t)  \!=\!\lambda^2 \Big[\int^{0}_{-t} \! ds {\cal P} {\cal L} (\tau) {\cal L}(s) {\cal P} \tilde{\xi}_{\rm SR}(\tau,t) \!-\! \!\int^{0}_{-t} ds {\cal P} {\cal L}(\tau) \nonumber \\
&&\big[\rho_{\rm SR}(t) {\cal L}(s) O_{1}(0)\big]\Big]\!-\!{\cal F}(\tau) {\cal Q} \Big[e^{-iH_{0}t}\rho_{\rm SR}(0) e^{iH_{0}t}O_{1}(0)\Big],
 \end{eqnarray}
where ${\cal F}(\tau)$ has $O(\lambda)$ and $O(\lambda^2)$ terms (See \cite{supp} for the details of the derivation).
Interestingly, under the assumption that $[O_1,S]=0$ i.e., $O_1$ being diagonal in the basis of $S$, and the strong Markovian approximation $\tau \gg \tau_B$, the inhomogeneous terms in Eq.~\eqref{eqnzm} vanishes and consequently the effective density matrix satisfies a rate equation of the form given in Eq.~\eqref{opd}. As a consequence, under the above conditions, the QRT relation emerges in the USC regime. 

We next consider a more general scenario where $O_{2}$ is not diagonal in the basis of $S$ but $O_1$ remains diagonal. In this case, the off-diagonal parts of the effective density matrix also contribute to the correlation function $\langle O_{1}(t)O_{2}(t+\tau)\rangle$, as follows from Eq.~\eqref{corr-gen}. Below, we present the recipe to compute the off-diagonal part of $\xi_{S}(\tau,t)$ upto the first order in $\lambda$, under the strong Markov approximation.

\subsubsection{Off-diagonal elements of the effective density matrix}
To obtain the off-diagonal elements of the effective density matrix  , we do not consider the dynamics of the effective density matrix but rather work directly with the definition of $\xi_{S}(\tau,t)$.
The off-diagonal part of $\xi_{S}(\tau,t)$ can be expressed as
\begin{align}\label{effofda}
 \xi^{nm}_{S}(\tau,t)&=\langle n| \xi_{S} (\tau,t)| m\rangle\nonumber\\
  &=\mybra{n} {\rm Tr}_{R}\Big[e^{-iH_{0}\tau}\big(\tilde{\xi}_{\rm SR}(\tau,t)\big)e^{iH_{0}\tau}\Big]\myket{m} \nonumber \\
  &=\mybra{n} {\rm Tr}_{R}\Big[e^{-iH_{0}\tau}\big({\cal Q} \,\tilde{\xi}_{\rm SR}(\tau,t)\big)e^{iH_{0}\tau}\Big]\myket{m} 
\end{align}
Note that, the above equation is exact. In the last line, the presence of ${\cal Q}$ gives the same result for the effective density matrix as ${\cal Q}= I- {\cal P}$ and the contribution from the ${\cal P}$ part vanishes. Our aim here is to find the $ \xi^{nm}_{S}(\tau,t)$ upto leading order in $\lambda$.  The formal expression of $ \xi^{nm}_{S}(\tau,t)$ under the strong Markov approximation is given as (please see Eq.~\eqref{o} in \cite{supp} for the details of the derivation)
\begin{eqnarray}
\label{offdm}
 \xi^{nm}_{S}(\tau,t)&=&-i\lambda
 J_{nm} \Big[ \tilde{\xi}^{m}_{S}(\tau,t)
\int^{\tau}_{0}d\tau' e^{-i(\tilde{\epsilon}_{n}-\tilde{\epsilon}_{m})\tau'}\zeta_{nm}(\tau')\nonumber\\
&-& \tilde{\xi}^{n}_{S}(\tau,t)
\int^{\tau}_{0}d\tau' e^{-i(\tilde{\epsilon}_{n}-\tilde{\epsilon}_{m})\tau'}\zeta^*_{mn}(\tau')\Big],
\end{eqnarray}
where $\zeta_{nm}(s) $ is the bath correlation function and is defined as $\zeta_{nm}(s) ={\rm Tr}\big[e^{iH^{m}_{R}s}e^{-iH^{n}_{R}s}\rho^{m}_{R}\big]$. It is worth noting that, we have expressed the off-diagonal part $ \xi^{nm}_{S}(\tau,t)$ in terms of the diagonal elements $ \xi^{n}_{S}(\tau,t)$. 
This result exhibits the same structural relationship as seen in the density matrix case, with off-diagonal elements expressed through diagonal ones \cite{supp}.
Thus the QRT, as presented in Eq.~\eqref{eqqrt} also holds in this case.  

Until now, we have considered the operator 
$O_{1}$ to be diagonal and discuss QRT. In the following, we relax this assumption and explore how QRT arises when $O_{1}$ is no longer diagonal.

\noindent {\it Case 2: Non-Diagonal  $O_{1}$ Operator.--} It is important to highlight that the homogeneous part of Eq.~\eqref{eqnzm} is independent of whether the operator $O_{1}$ is diagonal or not. However, the inhomogeneous part crucially depends on the same. We earlier showed in Eq.~\eqref{eqinhaa-main} that the inhomogeneous part contain only ${\cal O}(\lambda^2$) terms for diagonal $O_{1}$. For the non-diagonal $O_{1}$, the inhomogeneous part contain both ${\cal O}(\lambda)$ and ${\cal O}(\lambda^2)$ order terms. Since, this extra term is of ${\cal O}(\lambda)$, it will have a significant effect when $\tau$ is small. Below, we write the explicit expression of the  ${\cal O}(\lambda)$ inhomogeneous term for non-diagonal $O_1$ (See Eq.~\eqref{eqinha} in \cite{supp}) 
\begin{equation}
\!\!\!\!I(\tau)  {\cal Q} \tilde{\xi}_{\rm SR}(0,t)\!=\!\lambda {\cal P} {\cal L}(\tau) {\cal Q} \Bigg[\sum_{n} \!p_{n}(t)\myket{n}\mybra{n}O_{1}\otimes \rho^{n}_{R} \Bigg].
\label{inhom}
\end{equation}
By inserting the expression of ${\cal L}(\tau)$, Eq.~\eqref{inhom} can be expressed as  
\begin{align}\label{inhf}
   & I(\tau)\, {\cal Q}\, \tilde{\xi}_{\rm SR}(0,t)\!=-i\lambda  \sum_{n,m} \Big[ J_{nm} p_{m}(t)O^{mn}_{1}e^{i(\tilde{\epsilon}_{n}-\tilde{\epsilon}_{m})\tau}\zeta^*_{mn}(\tau)\nonumber\\
    &-J_{mn} p_{n}(t)O^{nm}_{1}e^{i(\tilde{\epsilon}_{m}-\tilde{\epsilon}_{n})\tau}\zeta^*_{nm}(\tau)\Big] \myket{n}\mybra{n}\otimes\rho^{n}_{R} 
\end{align}
which shows a non-trivial ${\cal O}(\lambda)$ contribution as long as $[O_1,S] \neq 0$. Note that such a contribution can drastically affect the validity of QRT for small $\tau$, unlike the standard weak-coupling scenario. However once $\tau \gg \tau_B$, the bath correlation function $\zeta_{mn}(\tau)$ decays rapidly and restores QRT as like Eq.~\eqref{eqnzm} without the inhomogeneous part. Note that this ${\cal O}(\lambda)$ correction term is diagonal in the $S$ basis. 

Next, we look at the off-diagonal part of $\xi_{\rm SR}(\tau,t) $ when $O_1$ is non-diagonal. We calculate $ \xi^{nm}_{S}(\tau,t)$ following the similar procedure  as done using Eq.~\eqref{effofda}. Now $O_1$ being  non-diagonal, $\xi^{nm}_{S}(\tau,t)$ has a contribution from the zeroth order in $\lambda$, which was absent when $O_{1}$  is diagonal. The zeroth order contribution is given as [see Eq.~\eqref{effofdaa} in \cite{supp}]  
\begin{align}
\!\!\!\!\!\xi^{nm}_{S}(\tau,t)=O^{nm}_{1}p_{n}(t) e^{-i(\tilde{\epsilon}_{n}-\tilde{\epsilon}_{m})\tau} \zeta^*_{mn}(\tau).
\end{align}
Note that for a given $t$, the decay time-scale of the above off-diagonal elements $\xi^{nm}_{S}$ is determined by the  strength of the system-bath interaction. Since this is strong, the corresponding decay occurs on a much shorter time scale compared to that of the diagonal elements, whose decay is instead governed by the small perturbative parameter in the  Hamiltonian $V$ in Eq.~\eqref{eq:V}. Our analysis thus reveals a novel feature-- the emergence of two distinct decay time scales in the correlation functions. These time scales arise from the contributions of the bare [$H_0$ in Eq.~\eqref{H0-dressed}] and interaction part [$V$ in Eq.~\eqref{eq:V}] of the total Hamiltonian. Notably, such behavior is absent in the weak system-bath coupling regime where relaxation time-scale is governed only by the weak system-bath coupling strength and is a characteristic feature of the strong system-bath coupling regime.

\vspace{0.3cm}
{\it Example: Dissipative Spin-Boson Model.--}
We now apply our method for computing the correlation function in a paradigmatic model, namely, the dissipative spin-boson model \cite{breuer, Carmichael,rivas2012open,Agarwalla_2017}. It consists of a two-level system coupled to a bosonic bath. We write the total Hamiltonian of the setup as
\begin{equation} 
\label{spm}
\begin{split}
H = \epsilon \sigma_{z}+\sum_{k} \Omega_{k} b_{k}^\dagger b_{k}+ \sigma_{x} \sum_{k} \alpha_{k} (  b_{k}^\dagger+ b_{k}),
\end{split}
\end{equation}
where 
$\sigma_{i}, i=x, y,z$ are the standard Pauli matrices. Note that the Pauli operator $\sigma_x$ is coupled with the bath degrees of freedom. Here, we focus on computing correlation functions that involve both the diagonal and the off-diagonal parts of $\xi_{S}(\tau,t)$.

\noindent {\it Example with diagonal $O_{1}$ and $O_{2}$.--} Let us first consider a two-point correlation function for $O_{1}$ which is diagonal in $\sigma_x$ basis. We consider $O_1=O_2=\myket{+}\mybra{+}$ and compute the correlation  $C_{++}(\tau,t)=\langle \myket{+}\mybra{+}(t)\myket{+}\mybra{+} (t+\tau) \rangle$, where, $\myket{\pm}$ represent the  eigenstates of the Pauli operator $\sigma_{x}$, and in  the computational basis they are given by $\myket{\pm}=(\myket{0}\pm\myket{1})/\sqrt{2} $. Thus, in terms of the effective density matrix  $ C_{++}(\tau,t)=\mybra{+}\xi_{S}(\tau,t)\myket{+}$.

\begin{figure}
    \centering
     \includegraphics[width=\columnwidth]{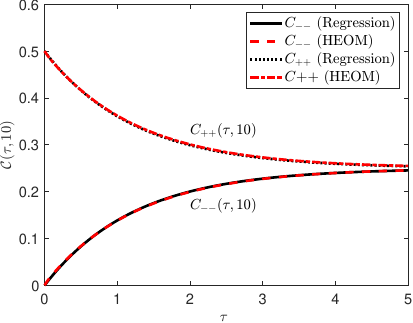}
    \caption{Plot for the correlation
    functions involving diagonal operators $C_{++}(\tau,t)$ and $C_{--}(\tau,t)$ as a function of $\tau$ for a fixed value of $t=10$ using the analytical approach and the numerically exact HEOM approach. For the $C_{++}(\tau,t)$ [$C_{--}(\tau,t)$], the black dotted [black solid] line represents the analytical  result that follows QRT, while the red dashed dot [dashed] line corresponds to the HEOM result. As the initial state we consider for the system $\rho_{S}(0)=\myket{+}\mybra{+}$. We take the  Drude-Lorentz spectral density for the bath $J(\Omega)=\frac{2 \Lambda\Delta \Omega}{\pi(\Omega^2 +\Delta^2)}$, with $\Delta$ being the width of the spectrum and $\Lambda$ is the coupling strength. Here we choose $\Lambda=100$, $\Delta=50$.  The inverse temperature of the bath $\beta=4.5 \times 10^{-3}$, and $\epsilon=10$.}
    \label{fig:C++}
\end{figure}

Introducing $q_{\pm}(\tau,t) =\mybra{\pm}\xi_{S}(\tau,t)\myket{\pm}$ and using the QRT, it is easy to show that 
\begin{align}
\label{dre}
   & \frac{d}{d\tau}q_{\pm}= \mp \gamma_{+}q_{+} \pm \gamma_{-}q_{-},
\end{align}
where $\gamma_{+}$ and $\gamma_{-}$ are the rates and related to the bath correlation function $\zeta_{-+}(\tau)={\rm Tr}\big[e^{iH^{-}_{R}\tau}e^{-iH^{+}_{R}\tau}\rho^{-}_{R}\big]$. 
The explicit form is given as
\begin{equation}
\!\!\!\!\gamma_{+}=\epsilon^2 \int^{\infty}_{-\infty}\! d\tau' e^{-i(\bar{\epsilon}_{-}-\bar{\epsilon}_{+})\tau'} \zeta_{-+}(\tau')\!=\!e^{-\beta (\bar{\epsilon}_{-}-\bar{\epsilon}_{+})} \gamma_{-}
\end{equation}
where  $\bar{\epsilon}$ is the  dressed onsite energy. For this model it is easy to show that $\bar{\epsilon}_{+} = \bar{\epsilon}_{-}$. Thus we note $\gamma_{+}=\gamma_{-}=\gamma$. Solving Eq.~\eqref{dre}, we obtain the following result
\begin{align}
\label{tpqrt}
  q_{+}(\tau,t)= p_{+}(t)+ \frac{p_{+}(t)}{2}\Big[e^{-2\gamma\tau}-1\Big],
\end{align}
where $p_{+}(t)=\mybra{+}\rho_{S}(t)\myket{+}$ represents the population in the $\myket{+}$. To find the population, one can use the QME in Eq.~\eqref{opd} and obtain 
\begin{align}\label{ddm}
   p_{+}(t)= p_{+}(0)+\frac{p_{+}(0)-p_{-}(0)}{2}[e^{-2\gamma t}-1].
\end{align}
As a result, we now have an explicit expression for the  $C_{++}(\tau,t)$ correlator [Eq.~\eqref{tpqrt}
and Eq.~\eqref{ddm}]. In Fig.~\ref{fig:C++}, we plot $C_{++}(\tau,t)$ as a function of $\tau$ for a fixed value of $t$ using our approach and also verify our results following the numerically exact hierarchical equations of motion (HEOM) method. As can be seen, the results from both approaches match very well. Moreover, for a particular $t$, the saturation time-scale of $C_{++}(\tau,t)$ is dictated by the inverse of the hopping strength $1/2\gamma$ with $\gamma \propto \epsilon^2$.  We also plot  $C_{--}(\tau,t)=\mybra{-}\xi_{S}(\tau,t)\myket{-}$ as a function of 
$\tau$.

\noindent {\it {Example with off-diagonal $O_{1}$ and $O_{2}$.--}}
We next consider an example where  
$O_{1}$ and $O_2$ both are not diagonal in the 
$\sigma_{x} $ basis. Specifically, we take $O_{1}=\big(\frac{1}{2} \myket{+}\mybra{+}+\myket{+}\mybra{-}\big)$ and $O_{2}=\big(\myket{+}\mybra{+}\, +\, \myket{-}\mybra{+}\big)$. In order to compute $ C_{+-}(\tau,t)=\langle \big(\frac{1}{2}\myket{+}\mybra{+}+\myket{+}\mybra{-}\big)(t)\big(\myket{+}\mybra{+}+\myket{-}\mybra{+}\big) (t+\tau) \rangle$ correlator, we need both diagonal and off-diagonal elements of the effective density matrix. More precisely,  $ C_{+-}(\tau,t)=\mybra{+}\xi_{S}(\tau,t)\myket{+}+\mybra{+}\xi_{S}(\tau,t)\myket{-}$. We compute this quantity following the method prescribed above. 

\begin{figure}
    \centering
    \includegraphics[width=\columnwidth]{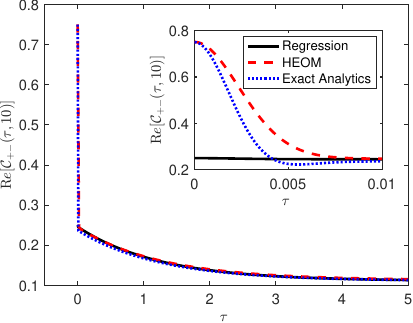}
    \caption{Plot for the real part of correlation function $C_{+-}(\tau,t)$ that involves off-diagonal $O_1$ operator, as a function of $\tau$ for a fixed value of $t$ using the  analytical result (blue dotted), prediction following QRT (black solid) and the HEOM (red dashed) approach. The plot clearly shows the existence of two different time scales. The inset shows the zoomed version of the short time dynamics for the correlator. The conventional QRT completely fails to predict the dynamics. The parameters used here are identical to that of Fig.~\ref{fig:C++}.}
    \label{fig:20}
\end{figure}

In Fig.~\ref{fig:20}, we plot $C_{+-}(\tau,t)$ that involves off-diagonal operators $O_1$ and $O_2$, as a function of $\tau$ for a fixed value of $t$. We compare the results obtained using our approach and from the HEOM. Since both $O_1$ and $O_2$ are off-diagonal operators, our analytical results predict the emergence of two distinct decay time scales in the correlator. The fast decay occurs on a time scale set by the inverse of the system-bath coupling strength, while the subsequent slow relaxation takes place on a time-scale determined by $1/\epsilon^2$. This is clearly demonstrated in Fig.~\ref{fig:20}. Furthermore, as shown in the inset, the conventional QRT completely fails to capture the initial rapid decay of the correlation function (black solid line), whereas our analytical result and HEOM both predict the fast decay. Note that, the small discrepancy between HEOM and the exact analytical results arises from differences in the choice of initial conditions for the bath. 


{\it Discussion.--}
\label{sec-V}
This work addresses the largely unexplored problem of two-point correlation functions in the ultra-strong system-bath coupling regime, an area of growing relevance in fields like open quantum systems, quantum thermodynamics, and quantum thermometry. While prior studies have extensively examined the dynamics of the density matrix using numerical methods such as HEOM, reaction coordinate, polaron transformation, and  TEMPO, correlation functions beyond weak coupling remain less understood.

Using the Nakajima-Zwanzig projection operator formalism, we analytically studied the dynamics of two-point correlation functions in this regime. A key question we addressed is the validity of the Quantum Regression Theorem (QRT) under the ultra-strong coupling. 
We found that in the USC regime, depending on the type of two-time correlator, there could be two distinct time-scales and the QRT emerges once the fastest time-scale has elapsed. For the diagonal class of operators, the validity of QRT for arbitrary correlation time $\tau$ is much better than the off-diagonal case. This is because of the emergence of two distinct decay time scales—arising from the bare and interaction parts of the Hamiltonian. This behavior is absent in the weak-coupling limit and underscores the rich dynamics induced by strong coupling. Finally, we benchmarked our analytical results against numerically exact HEOM simulations and observed excellent agreement, supporting the accuracy and utility of our approach. 
It would be interesting to understand the long-time thermodynamically consistent thermalization behavior such as the emergence of Kubo-Martin-Schwinger condition for such two-point correlators in the USC limit along with impact of non-Markovian corrections.

\section*{Acknowledgements} 
SK acknowledges the CSIR fellowship with Grant Number 09/0936(11643)/2021-EMR-I. BKA acknowledges CRG Grant No. CRG/2023/003377 from Science and Engineering Research Board (SERB), Government of India. BKA and SK would like to thank Sachin Jain for detailed  discussions. SK thanks Archak Purkayastha and BKA thanks Gerlad Fux for many insightful discussion related to this project.  BKA acknowledges the hospitality and support of the International Centre of Theoretical Physics (ICTP), Italy under the associateship program.

\bibliography{reference.bib}

\onecolumngrid

\setcounter{equation}{0}
\setcounter{figure}{0}
\setcounter{section}{0}
\renewcommand{\theequation}{S\arabic{equation}}
\renewcommand{\thefigure}{S\arabic{figure}}

\begin{center}
{\textbf{\underline{Supplementary Material}}}
\end{center}

\section{Review of Nakajima-Zwanzig projection operator method}
In this section, we briefly review the Nakajima-Zwanzig projection operator formalism. We begin by decomposing the total system-bath Hamiltonian  as
$H=H_{S}+H_{R}+ H_{S R}=H_{0}+\lambda V$, where $H_0$ denotes the unperturbed part of the Hamiltonian that includes the system-bath interaction Hamiltonian and $V$ represents the perturbative part of the total Hamiltonian $H$.
In the interaction picture with respect to $H_0$, the time evolution of the full density matrix $\rho_{\rm SR}$ is governed by
 \begin{align}
     \dot{\tilde{\rho}}_{\rm SR}(t)=-i\lambda\,[\tilde{V}(t),\tilde{\rho}_{\rm SR}(t)]=\lambda\, {\cal L}(t) \tilde{\rho}_{\rm SR}(t),
     \label{full-den-interaction}
 \end{align}
 where the interaction picture operators are defined as 
 \begin{align}
    & \tilde{\rho}_{\rm SR}(t)=e^{iH_{0}t}\; \rho_{\rm SR}(t) \;e^{-iH_{0}t},\nonumber\\
     & \tilde{V}(t)=e^{iH_{0}t}\; V\;e^{-iH_{0}t}, 
 \end{align}
and the time-dependent superoperator ${\cal L}(t)$ acts as ${\cal L}(t) \, * = -i \big[\tilde{V}(t), *\big]$. We next introduce two complementary projection superoperators ${\cal P}$ and ${\cal Q}$ that satisfy the following properties:
\begin{align}\label{pop}
   & {\cal P}+{\cal Q}=I,\hspace{.5cm} {\cal P}^2={\cal P},\nonumber\\
    &{\cal Q}^2={\cal Q}, \hspace{.3 cm} {\cal P} {\cal Q}= {\cal Q} {\cal P}=0.
\end{align}
Using this decomposition, the exact time-evolution for $\tilde{\rho}_{\rm SR}(t)$ can be split as follows \cite{breuer}
\begin{align}\label{pod}
   & {\cal P} \dot{\tilde{\rho}}_{\rm SR}(t)=\lambda \, {\cal P} \, {\cal L}(t) {\cal P}\tilde{\rho}_{\rm SR}(t)+\lambda\, {\cal P} \, {\cal L}(t) \, {\cal Q} \tilde{\rho}_{\rm SR}(t),\nonumber\\
    & {\cal Q} \tilde{\rho}_{\rm SR}(t)\!=\! {\cal G}(t,0) {\cal Q} \tilde{\rho}_{\rm SR}(0)\!+\!\lambda \int^{t}_{0} \! ds \, {\cal G}(t,s)\, {\cal Q} \, {\cal L}(s) {\cal P} \tilde{\rho}_{\rm SR}(s)
\end{align}
where we have introduced the propagator $ {\cal G}(t,s)=e^{\lambda \int^{t}_{s} ds' {\cal Q} {\cal L}(s')}$. This equation is in fact known as the Nakajima—Zwanzig equation \cite{breuer}. So far this is an exact equation for the density matrix. Now if $ {\cal P} {\cal L}(t){\cal P}=0$ then ${\cal Q} \tilde{\rho}_{\rm SR}(t) $ up to first order in $\lambda$ can be written as 
\begin{align}\label{offdd}
    {\cal Q} \tilde{\rho}_{\rm SR}(t)=&{\cal Q} \, \tilde{\rho}_{\rm SR}(0)+\lambda \int^{t}_{0} ds {\cal Q} {\cal L}(t-s) {\cal Q} \, \tilde{\rho}_{\rm SR}(0)\nonumber\\
    &+\lambda \int^{t}_{0} ds {\cal L}(t-s) {\cal P} \, \tilde{\rho}_{\rm SR}(t-s)
\end{align}
By substituting the above expression in Eq.~\eqref{pod}, we receive the master equation for ${\cal P} \tilde{\rho}_{\rm SR}(t) $ with inhomogeneous terms and is correct up to $O(\lambda^2)$ 
\begin{align}
\label{nzf}
   & {\cal P} \dot{\tilde{\rho}}_{\rm SR}(t)=\lambda^2 \int^{t}_{0} ds {\cal P} {\cal L}(t){\cal L}(t-s)\big[{\cal P} \tilde{\rho}_{\rm SR}(t)+{\cal Q} \tilde{\rho}_{\rm SR}(0)\big]+ \lambda {\cal P} {\cal L}(t) {\cal Q} \rho_{\rm SR}(0)
\end{align}
where note that in the above equation we replace $\tilde{\rho}_{\rm SR}(t-s)$ by $\tilde{\rho}_{\rm SR}(t)$ which is justified as we are interested up to $O(\lambda^2)$.

 \section{Dynamics of two-point correlation functions at the ultra-strong coupling regime}
 Let us consider a system of interest that interacts with a bosonic thermal bath. The Hamiltonian of the combined system can be expressed as 
\begin{equation} \label{eq1nmaa}
		\begin{split}
			H = H_{S}+\sum_{k} \Omega_{k} b_{k}^\dagger b_{k}+ S \sum_{k} \alpha_{k} (  b_{k}^\dagger+ b_{k})= H_{0}+ \lambda V\;,
		\end{split}
\end{equation}
 where $H_{0}=H^{d}_{S}+\sum_{k} \Omega_{k} b_{k}^\dagger b_{k}+ S \sum_{k} \alpha_{k} (  b_{k}^\dagger+ b_{k})$ and $V=H^{c}_{S}$. Here, $H^{d}_{S}$ and $H^{c}_{S}$ represent the diagonal and off-diagonal components of the system Hamiltonian $H_{S}$ in the eigenbasis of the Hermitian system operator $S$, respectively.
In our case, the action of the projection operator ${\cal P}$  is defined as $ {\cal P} X_{\rm SR}=\sum_{n}\mybra{n}{\rm Tr}_{R}[X_{\rm SR}]\myket{n} \myket{n}\mybra{n}\otimes \rho^{n}_{R}$ and ${\cal Q}=I-{\cal P}$ with 
\begin{equation}
\label{rhoR-n}
\rho^{n}_{R}=\frac{e^{-\beta H^{n}_{R}}}{{\rm Tr}[e^{-\beta H^{n}_{R}}]} =\frac{e^{-\beta (\mybra{n}\sum_{k} \Omega_{k} b^{n \dagger}_{k} b^{n}_{k} \myket{n})}}{{\rm Tr}[e^{-\beta (\mybra{n}\sum_{k} \Omega_{k} b^{n \dagger}_{k} b^{n}_{k} \myket{n})}]},
\end{equation}
where $b^{n}_{k}$ are the displaced bath annihilation operators, defined as $b^{n}_{k}=b_{k}+\frac{\theta_{n}\alpha_{k}}{\Omega_{k}}\myket{n} \mybra{n}$.  Note that, the projection operators ${\cal P}$ and ${\cal Q}$ obey the properties defined in Eq.~\eqref{pop}.

The diagonal and off-diagonal parts of the system Hamiltonian $H_{S}$ can be written as $H^{d}_{S}=\sum_{n}\epsilon_{n}\myket{n}\mybra{n}$ and $V=\sum_{n\neq m}J_{nm}\myket{n}\mybra{m}$. In the interaction picture, $V$ is given by
\begin{align}\label{vi}
    \tilde{V}(t)=\sum_{n\neq m} J_{nm}e^{i(\tilde{\epsilon}_{n}-\tilde{\epsilon}_{m})t}e^{iH^{n}_{R}t}\myket{n}\mybra{m}e^{-iH^{m}_{R}t},
\end{align}
where recall that $ \tilde{\epsilon}_{n}=\epsilon_{n}-\sum_{k}\frac{\alpha_{k}^{2}}{\Omega_{k}}\theta_{n}^{2}=\epsilon_{n}-\delta\epsilon_{n}$ is the dressed onsite energy.
We are interested to compute the two-point correlation function of the form  $C(\tau,t)=\langle O_{1}(t)O_{2}(t+\tau) \rangle$. This can be written as 
\begin{align}
     \langle O_{1}(t)O_{2}(t+\tau) \rangle= {\rm Tr}_{S}\big[O_{2}(0)\xi_{S}(\tau,t)\big],
 \end{align}
where $\xi_{S}(\tau,t)= {\rm Tr}_{R}[e^{-iH\tau} \rho_{\rm SR}(t) O_{1}(0)e^{iH\tau}]$. In the interaction picture the effective full density matrix $\xi_{\rm SR}(\tau,t)=e^{-iH\tau} \rho_{\rm SR}(t)\, O_{1}(0)\,e^{iH\tau}$
satisfies the same equation as the full density matrix in Eq.~\eqref{full-den-interaction} i.e.,
\begin{align}
     \frac{d}{d\tau} {\tilde{\xi}}_{\rm SR}(\tau,t)=\lambda \, {\cal L}(\tau) \, \tilde{\xi}_{\rm SR}(\tau,t).
 \end{align}
Employing the Nakajima-Zwanzing projection operator method, as described in the previous section, one can  show the following [similar to Eq.~\eqref{nzf}]
\begin{align}
\frac{d}{d\tau} {\cal P} {\tilde{\xi}}_{\rm SR}(\tau,t) &= \lambda^2 \! \int^{\tau}_{0} ds {\cal P} {\cal L}(\tau) {\cal L}(\tau-s)\Big[{\cal P} \, \tilde{\xi}_{\rm SR}(\tau,t) + \nonumber \\
& {\cal Q} \, \tilde{\xi}_{\rm SR}(0,t)\Big]+ \lambda {\cal P} {\cal L}(\tau) {\cal Q} \tilde \xi_{\rm SR}(0,t).
\end{align}
From the above equation, it is clear that the dynamical equation governing the two-point correlation function also contains an inhomogeneous part. The above equation can be re-written as a sum of homogeneous and an inhomogeneous term as
\begin{equation}
\label{eqnz}
\frac{d}{d\tau} {\cal P} \tilde{\xi}_{\rm SR}(\tau,t)  =\lambda^2 \!\int^{\tau}_{0} ds  {\cal P} {\cal L}(\tau) {\cal L}(\tau-s) {\cal P} \tilde{\xi}_{\rm SR}(\tau,t)+{I(\tau) {\cal Q}\, \tilde{\xi}_{\rm SR}(0,t)},
 \end{equation}
where the inhomogeneous part in Eq.~\eqref{eqnz} is given by
\begin{equation}
    \label{inh1}
I(\tau) {\cal Q} \tilde{\xi}_{\rm SR}(0,t)=\Big[\lambda {\cal P} {\cal L}(\tau)\Big(\mathbb{I}+\lambda\int^{\tau}_{0} ds {\cal L}(\tau-s)\Big)\Big]{\cal Q} \tilde{\xi}_{\rm SR}(0,t)
\end{equation}
We want to now expand the inhomogeneous term upto $\mathcal{O}(\lambda^2)$. Note that, the action of ${\cal Q} $ operator on $\tilde{\xi}_{\rm SR}(0,t)=\rho_{\rm SR}(t) O_{1}(0)$ is non-zero and upto first order in $\lambda$ is given by
\begin{align}
\label{qxi0}
{\cal Q}\, \tilde{\xi}_{SR}(0,t) = {\cal Q}\Big[e^{-iH_{0}t}\rho_{\rm SR}(0) e^{iH_{0}t}O_{1}(0)\Big] + \lambda \int^{0}_{-t} ds \,  {\cal Q} \, {\cal L}(s) \tilde{\xi}_{\rm SR}(0,t) - \lambda \int^{0}_{-t} ds  {\cal Q} \, \Big[\rho_{\rm SR}(t) {\cal L}(s)O_{1}(0)\Big]+O(\lambda^2)\;.
 \end{align}
So finally, the inhomogeneous part defined in Eq.~\eqref{inh1} upto ${\cal O}(\lambda^2)$ can be expressed as
\begin{align}
\label{eqinha}
   I(\tau) \, Q \, \tilde{\xi}_{\rm SR}(0,t)&=\lambda^2\int^{0}_{-t} ds \, {\cal P} {\cal L}(\tau) {\cal L}(s) ({\cal P}+ {\cal Q})  \tilde{\xi}_{\rm SR}(\tau,t)-\lambda^2 \int^{0}_{-t} ds \, {\cal P} {\cal L}(\tau)\, \Big[\rho_{SR}(t){\cal L}(s)O_{1}(0)\Big]\nonumber \\
   &+\Bigg(\lambda P {\cal L}(\tau)+\lambda^2\int^{\tau}_{0}ds\; {\cal P} {\cal L}(\tau) {\cal L}(\tau-s) \Bigg) {\cal Q} \Big[e^{-iH_{0}t}\rho_{\rm SR}(0) e^{iH_{0}t}O_{1}(0)\Big]
 \end{align}

\section{Additional details for the derivation of the QRT}
\label{sec:appC}
In the weak coupling regime, the QRT is a widely used tool for computing correlation functions. Briefly,
the QRT states that the knowledge of the time evolution of a single-point function (density matrix) is sufficient to determine
the time evolution of two-point correlation
functions. More explicitly, let’s assume that the density matrix of the system, $\rho_{S} (t)$, at any time is given by 
\begin{equation}
\label{comaa}
	\rho_{S} (t)=\Phi(t,0)\,\rho_{S}(0).
\end{equation}
Then the QRT for the two functions read as \cite{breuer}
\begin{align}\label{eqqrta}
	\langle O_{1}(t)O_{2}(t+\tau)\rangle&\!=\! {\rm Tr}_{S}\big[O_{2}\,\Phi(t+\tau,t)\,[\xi_{S}(0,t)]\big].
\end{align}
Note that, to derive the QRT in
Eq. \eqref{eqqrta}, apart from assuming weak coupling between the
system and bath, the strong Markovian limit i.e., $ \tau\gg\tau_{B}$ is assumed, where $\tau_B$ is the characteristic decay time-scale for the bath correlation function. Below, we discuss how QRT emerges in the ultra strong coupling regime.

\subsection{Dynamics of the reduced density matrix}
We consider that the the system-bath state at initial time $t=0$ to be of the form $\rho_{\rm SR}(0)=\sum_{n}p_{n}\myket{n}\mybra{n}\otimes \rho^{n}_{R}$. As a result, the initial state is invariant under the action of the projection operator ${\cal P}$ i.e., ${\cal P} \rho_{\rm SR}(0)=\rho_{\rm SR}(0)$ and ${\cal Q} \rho_{\rm SR}(0)=0$. As a consequence, the inhomogeneous part of the master equation in Eq.~\eqref{nzf} vanishes, and we receive the following homogeneous quantum master equation 
\begin{align}
\label{standard-QME}
   & {\cal P} \dot{\tilde{\rho}}_{\rm SR}(t)=\lambda^2 \int^{t}_{0} ds \, {\cal P} {\cal L}(t) \, {\cal L}(t-s)\, {\cal P} \tilde{\rho}_{\rm SR}(t).
\end{align}
One can further simplify the above equation and obtain a Pauli rate-type equation in the basis of $S$ and is given as
\begin{align}
   &  \dot{p}_{n}(t)=\lambda^2 \sum_{m\neq n}[\gamma_{nm}(t) \, p_{m}(t)-\gamma_{mn}(t) \, p_{n}(t)],
\end{align}
where the rate constant $\gamma_{nm}(t)$ is defined as
\begin{align}
\label{rate-USC}
   & \gamma_{nm}(t)=2 |J_{nm}|^2 \int^{t}_{0} ds\; e^{-i(\tilde{\epsilon}_{n}-\tilde{\epsilon}_{m})s} \, {\rm Tr}\big[e^{iH^{m}_{R}s}e^{-iH^{n}_{R}s}\rho^{m}_{R}\big],
\end{align}
where recall that $\rho^{m}_{R}$ is given in Eq.~\eqref{rhoR-n}. If we assume that the bath correlation function decays rapidly with certain characteristic time scale $\tau_{B}$, then under the Markov approximation, i.e., $t\gg \tau_{B}$, the upper limit of the integration in Eq.~\eqref{rate-USC} can be extended to $\infty$. As a result, the rate constant becomes independent of time and one receives Eq.~\eqref{opd} of the main text.

It can be shown that the off-diagonal part of the density matrix can be expressed in terms of its diagonal elements [see Ref.\cite{PhysRevA.106.042209} for the details of the derivation]. Using Eq.\eqref{offdd}, we explicitly obtain the following relation for the density matrix  
\begin{align}\label{appeffofda}
 \rho^{nm}_{S}(t)&=\langle n| \rho_{S} (t)| m\rangle \nonumber \\
  &=-i\lambda
 J_{nm} \Big[ \tilde{\rho}^{m}_{S}(t)
\int^{t}_{0}d\tau' e^{-i(\tilde{\epsilon}_{n}-\tilde{\epsilon}_{m})\tau'}\zeta_{nm}(\tau')- \tilde{\rho}^{n}_{S}(t)
\int^{t}_{0}d\tau' e^{-i(\tilde{\epsilon}_{n}-\tilde{\epsilon}_{m})\tau'}\zeta^*_{mn}(\tau')\Big]
\end{align}

\subsection{Dynamics of the effective density matrix}
The dynamical equation of the two-point function i.e., effective density matrix, is given in Eq.~\eqref{eqnz}, and it contains an inhomogeneous term, $I(\tau) {\cal Q} \tilde{\xi}_{\rm SR}(0,t)$, which is given in Eq.~\eqref{eqinha}.  If the operator $O_{1}$ commutes with the $S$ operator i.e., $[S, O_1]=0$, the last term in the inhomogeneous part given in Eq.~\eqref{eqinha} vanishes. Then Eq.~\eqref{eqinha} reduces to
\begin{align}
\label{eqinhaa}
   &I(\tau) {\cal Q} \tilde{\xi}_{\rm SR}(0,t)=\lambda^2 \int^{0}_{-t} ds {\cal P} {\cal L} (\tau) {\cal L}(s) {\cal P} \tilde{\xi}_{\rm SR}(\tau,t)-\lambda^2 \int^{0}_{-t} ds  {\cal P} {\cal L}(\tau)\big[\rho_{\rm SR}(t) {\cal L}(s) O_{1}(0)\big].
 \end{align}
Under the strong Markovian approximation i.e., $\tau\gg \tau_{B}$, the above two inhomogeneous terms will also vanish (since, they are proportional to the bath correlation function), produces a homogeneous master equation for the effective density matrix i.e.,
\begin{equation}
\label{eqnz-1}
\frac{d}{d\tau} {\cal P} \tilde{\xi}_{\rm SR}(\tau,t)  =\lambda^2 \!\int^{\tau}_{0} ds  {\cal P} {\cal L}(\tau) {\cal L}(\tau-s) {\cal P} \tilde{\xi}_{\rm SR}(\tau,t).
\end{equation}
This equation is identical to the QME in Eq.~\eqref{standard-QME}.

We are now going to calculate the off-diagonal part of the effective density matrix.
Here also we assume that $O_{1}$ is diagonal in the eigenbasis of the system operator $S$ as we did in the previous derivation. The off-diagonal part of $\xi_{S}(\tau,t)$ can be expressed as
 \begin{align}\label{aappeffofda}
 \xi^{nm}_{S}(\tau,t)&=\langle n| \xi_{S} (\tau,t)| m\rangle\nonumber\\
  &=\mybra{n} {\rm Tr}_{R}\Big[e^{-iH_{0}\tau}\big({\cal Q} \tilde{\xi}_{\rm SR}(\tau,t)\big)e^{iH_{0}\tau}\Big]\myket{m}. 
\end{align}
Note that, the above equation is exact. However, we want to find the off-diagonal part that is correct only upto $\lambda$ order.  Using Eq.~\eqref{offdd}, we can write the off-diagonal part upto $\lambda$ order as
\begin{align}\label{}
 \xi^{nm}_{S}(\tau,t)=\mybra{n} {\rm Tr}_{R}\Big[e^{-iH_{0}\tau}\big({\cal Q} \, \tilde{\xi}_{\rm SR}(0,t)+\lambda \int^{\tau}_{0} d\tau' {\cal L}(\tau-\tau') {\cal P} \, \tilde{\xi}_{\rm SR}(\tau-\tau',t)\big)e^{iH_{0}\tau}\Big]\myket{m}.
\end{align}
Using Eq.~\eqref{qxi0}, we can show that the first term in the above equation, i.e.,  $\mybra{n} {\rm Tr}_{R}\Big[e^{-iH_{0}\tau}{\cal Q} \, \tilde{\xi}_{\rm SR}(0,t)e^{iH_{0}\tau}\Big]\myket{m}$ vanishes under the strong Markov approximation ($\tau \gg \tau_{B}$) and only the second term survives which is given by
\begin{align}
\label{o}
 \xi^{nm}_{S}(\tau,t)=&-i\lambda
 J_{nm} \Big[ \tilde{\xi}^{m}_{S}(\tau,t)
\int^{\tau}_{0}d\tau' e^{-i(\tilde{\epsilon}_{n}-\tilde{\epsilon}_{m})\tau'}\zeta_{nm}(\tau')\nonumber\\
&\hspace{1.7 cm}- \tilde{\xi}^{n}_{S}(\tau,t)
\int^{\tau}_{0}d\tau' e^{-i(\tilde{\epsilon}_{n}-\tilde{\epsilon}_{m})\tau'}\zeta^*_{mn}(\tau')\Big].
\end{align}
This reproduces the Eq.~\eqref{offdm} which is presented in the main text. This result exhibits the same structural relationship as seen in the density matrix case [refer to Eq.\eqref{appeffofda}], with off-diagonal elements expressed through diagonal ones.

\section{Additional details on two-time correlation  function}
\label{appd}
As we mentioned above, we have assumed $O_{1}$ commutes with the $S$ operator. However, in this subsection, we are going to get rid of this assumption. We want to now calculate the most general two-point correlation function of the form, $C(\tau,t)=\langle O_{1}(t)O_{2}(t+\tau) \rangle $, where both $O_{1}$ and $O_{2}$
can be any arbitrary system operator. Note that the homogeneous part of Eq.\eqref{eqnz} is independent of whether the operator $O_{1}$ is diagonal or not. However, the inhomogeneous part crucially depends on the same. In the previous section, we have shown that the inhomogeneous part contain only  $\lambda^2$ order terms for diagonal $O_{1}$. For the non-diagonal $O_{1}$, the inhomogeneous part contain both $\lambda$ and $\lambda^2$ order terms. Since, this extra term is of the order of $\lambda$, it can have a significant effect for small $\tau$. Below, we write the explicit expression of the inhomogeneous term [please see Eq.\eqref{eqinha}]
\begin{align}
\label{inhoma}
    I(\tau) \, {\cal Q} \, \tilde{\xi}_{\rm SR}(0,t)=\lambda {\cal P} {\cal L}(\tau) {\cal Q} \Big[\sum_{n}p_{n}(t)\myket{n}\mybra{n}O_{1}(0)\otimes\rho^{n}_{R} \Big],
\end{align}
where recall that if $O_1$ is diagonal in $S$ basis, this term vanishes.  Following Eq.\eqref{vi}, by inserting the expression of ${\cal L}(\tau)$ in the above equation, we write  Eq.~\eqref{inhoma} as
\begin{align}\label{inhaa}
   I(\tau) \, {\cal Q} \, \tilde{\xi}_{\rm SR}(0,t)=&-i\lambda  {\cal P}  \Big[\sum_{i,j}  J_{ij} p_{j}(t)e^{i(\tilde{\epsilon}_{i}-\tilde{\epsilon}_{j})\tau}\myket{i}\mybra{j}O_{1}e^{iH^{i}_{B}\tau}e^{-iH^{j}_{B}\tau}\rho^{j}_{R}\nonumber\\
    &\hspace{2.5 cm}-\sum_{i,j,n}  J_{ij} p_{n}(t)e^{i(\tilde{\epsilon}_{i}-\tilde{\epsilon}_{j})\tau}O^{ni}_{1}\myket{n}\mybra{j}\rho^{n}_{R}e^{iH^{i}_{B}\tau}e^{-iH^{j}_{B}\tau}\Big]\nonumber\\
    &=-i\lambda  \sum_{n,m} \Big[ J_{nm} p_{m}(t)O^{mn}_{1}e^{i(\tilde{\epsilon}_{n}-\tilde{\epsilon}_{m})\tau}\zeta^*_{mn}(\tau)\nonumber\\
    &\hspace{2.5 cm}-J_{mn} p_{n}(t)O^{nm}_{1}e^{i(\tilde{\epsilon}_{m}-\tilde{\epsilon}_{n})\tau}\zeta^*_{nm}(\tau)\Big] \myket{n}\mybra{n}\otimes\rho^{n}_{R}.
\end{align}
Hence, we derived the Eq.~\eqref{inhf} presented in the main text.

We want to now calculate the off-diagonal elements of the effective density matrix i.e., $\xi^{nm}_{S}(\tau,t)$, where $O_{1}$ can be any arbitrary system operator. We have computed the off-diagonal part of $\xi_{S}(\tau,t)$, and it has a contribution from the zeroth order in $\lambda$, which was absent in the previous case where $O_{1}$  is diagonal. The zeroth-order contribution is given by [refer to Eq.\eqref{effofda}]
\begin{align}
\label{effofdaa}
 \xi^{nm}_{S}(\tau,t)&=\mybra{n} {\rm Tr}_{R}\Big[e^{-iH_{0}\tau}Q\big[\sum_{k}p_{k}(t)\myket{k}\mybra{k}O_{1}(0)\otimes\rho^{k}_{R} \big]e^{iH_{0}\tau}\Big]\myket{m}\nonumber\\
 &=O^{nm}_{1}p_{n}(t) e^{-i(\tilde{\epsilon}_{n}-\tilde{\epsilon}_{m})\tau} {\rm Tr} [e^{iH^{m}_{R}\tau}e^{-iH^{n}_{R}\tau}\rho^{n}_{R}].
\end{align}
Note that, for a given $t$, the decay time scale of the above off-diagonal part is dictated by the parameter related to the strength of the system-bath interaction. Since, the strength of the system-bath interaction is large, the decay time scale is sufficiently small compared to the decay time scale of the diagonal part, dictated by the small parameter in the perturbative Hamiltonian $V$.

\end{document}